# On-demand control of photon echoes far exceeding the spin coherence constraint via coherence swapping between optical and spin transitions

Byoung S. Ham[*]
Center for Photon Information Processing, School of Electrical Engineering,
Inha University, 253 Yonghyun-dong, Nam-gu, Incheon 402-751, S. Korea
[*]bham@inha.ac.kr

Using on-demand coherence conversion via optical locking in conventional stimulated (three-pulse) photon echoes, ultralong, ultraefficient photon storage has been observed in a rare-earth $Pr^{3+}$ doped $Y_2SiO_5$, where storage time is several orders of magnitude longer and retrieval efficiency is enhanced by fifty times. Compared with the same method applied to two-pulse photon echoes for a rephasing halt, use of spectral grating in the stimulated photon echoes offers a phase-independent coherence conversion between optical and spin states. Eventually an ultralong spin population decay-dependent photon storage protocol is achieved by applying optical locking to a spectral grating in conventional stimulated photon echoes. The present breakthrough in both photon storage time and retrieval efficiency opens a door to long-distance quantum communications utilizing quantum repeaters.

Photon storage time far exceeding the critical constraint of optical phase decay time has been observed by adapting an optical locking method[1,2] to stimulated (three pulse) photon echoes[3], where photon echoes have intrinsic spatiotemporal multimode properties[1-12]. The dramatic extension of photon storage time has been achieved via population transfer-based coherence conversion between optical and spin states using the optical locking technique. Here we demonstrate photon storage nearly five orders of magnitude longer than that based on a rephasing halt such as in atomic frequency comb (AFC) echoes[11]. Moreover, we demonstrate echo retrieval efficiency nearly two orders of magnitude higher by adapting a phase conjugate scheme to solve the conventional echo reabsorption problem in a forward propagation scheme[4]. The observed photon storage time can be extended up to several hours keeping a near perfect retrieval efficiency at cryogenic temperatures[13]. Although the experimental demonstration was performed using classical light, the storage mechanism should be satisfied with quantum optical light owing to quantum mechanical properties of photon echoes[1-12]. This breakthrough in photon storage time and retrieval efficiency sheds light on intercontinental quantum communications for such as quantum cryptography, which rely on quantum repeaters[14].

Over the last decade, photon echoes[15] have been studied for multimode quantum memory applications based on reversible inhomogeneous broadening in order to increase echo efficiency[4-6], remove spontaneous emission noise[7-10], and extend photon storage time[11]. Although gradient echoes[8-10] and AFC[7,11] have successfully removed spontaneous emission noise caused by excited atoms in two-pulse photon echoes, these techniques still limit the photon storage time to the optical phase-decay constraint. To extend photon storage time, an optical locking technique has been suggested for stimulated on-resonance Raman echoes[16] to solve population decay-dependent coherence loss, where storage time extended by several orders of magnitude can be expected[2]. Here, we directly apply the optical locking method to stimulated photon echoes and experimentally demonstrate photon storage time extended by several orders of magnitude. We have also proved the theory of backward scheme of photon echoes to solve the intrinsic echo reabsorption dilemma in two-pulse photon echoes[4]. The observed photon storage time is extended up to spin population decay time, far exceeding the critical constraint of spin phase decay time in quantum memory protocols such as slow-light-based quantum memories[12,17-19], off-resonant Raman methods[20,21], and spin echoes in nitrogen vacancy diamonds[22]. By comparison, photon storage time in most modified photon echo methods is much shorter limited to optical phase constraint[4-11]. Although optical locking for a rephasing halt in two-pulse photon echoes has been demonstrated for photon storage extension[23], the extension factor is less than 10%, limited by the inverse of spin inhomogeneous broadening as also demonstrated in AFC echoes[11].

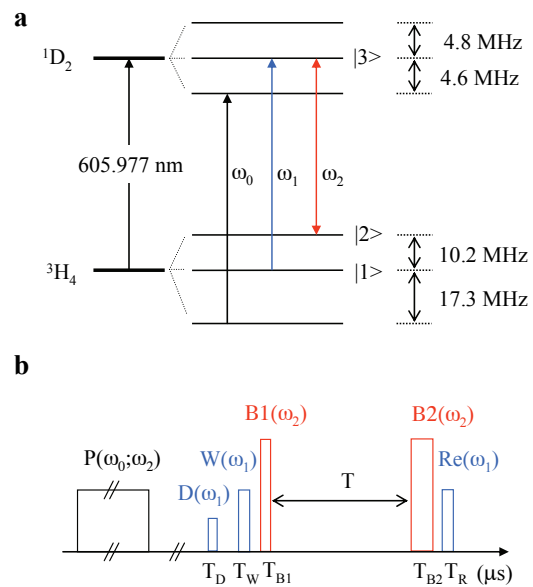

**Figure 1| Schematic of optical locking. a,** Energy level diagram of Pr:YSO for optically locked echoes. **b,** Light pulse sequence of (**a**).

Spontaneous emission noise caused by an optical rephasing pulse-induced population inversion is inevitable in conventional photon echoes. However, contrary to the harsh criticism raised over three-pulse (or stimulated) photon echoes[24], the effective number of atoms contributing to the quantum noise should be less than 0.03 out of $\sim 10^{18}$ excited

Pr$^{3+}$ ions in practical conditions of utilizing a full bandwidth of optical inhomogeneous width, which is negligibly small with regard to altering the quantum fidelity: see Fig. S1 in Supplementary Information. Here providing higher bandwidth of quantum memories is definitely desirable for practical quantum information processing as demonstrated recently with ns data pulses[21]. Therefore, the spontaneous emission noise in photon echoes using rare-earth doped solids cannot be a critical issue for quantum memory applications.

Figures 1a and 1b show the energy level scheme and pulse sequence of the present optically locked echoes, where these echoes stand for stimulated photon echoes controlled by the optical locking technique to avoid optical population decay-caused coherence loss. Optical locking is achieved by the deshelving optical pulse pair B1 and B2 at a frequency of $\omega_2$. To satisfy the phase recovery condition in an optically dense medium, B1 and B2 must satisfy the following conditions[1,23,25]:

$$\Phi_{B1}+\Phi_{B2}=4n\pi, \qquad (1)$$

$$\Phi_{B1}=(2n-1)\pi, \qquad (2)$$

where $\Phi$ is the pulse area, and n is an integer.

Figure 2 presents the coherence-population conversion process between off-diagonal elements $\rho_{13}$ (coherence) and diagonal elements $\rho_{11}$ (population), where $\rho_{33}=1-\rho_{11}$, in a two-level system composed of states |1> and |3>. In an inhomogeneously broadened optical system neglecting all decay parameters, a two-pulse photon echo is calculated in Figs. 2a~2d. The pulse area of D is set at $\pi/2$ for maximum coherence excitation. As shown in Fig. 2a, by the rephasing pulse R at t=20.0 μs, the system coherence ($i\rho_{13}$) shows optical phase reversal resulting in a photon echo at t=35 μs.

We now investigate the function of R for the coherence-population conversion. By the atom detuning Δ–dependent phase evolution [exp(−iΔt)], D-excited coherence $\rho_{13}$ results in a phase grating (see the top panels of Figs. 2c and 2d). Here $Re\rho_{13}$ has a $\pi/4$ phase shift against $Im\rho_{13}$ (not shown). When the rephasing pulse R comes in at t=20.0 μs, the phase grating $\rho_{13}$ starts to transfer into the population grating $\rho_{11}$ (see the middle panel of Fig. 2c). At the midpoint of pulse duration of R at t=20.1 μs for a $\pi/2$ pulse area (for example at the end of W in Fig. 1b), the coherence conversion from $\rho_{13}$ to $\rho_{11}$ is completed (see the middle panel of Fig. 2d): No phase grating ($i\rho_{13}$) remains at this time. At the end of R at t=20.2 μs for a $2\pi$ pulse area, the conversion process is completely reversed, where the initial phase grating is regained but with a $\pi$ phase shift (see the bottom panels of Figs. 2c and 2d). Here we note that the phase information of D is completely transferred into the population information $\rho_{11}$ by the half duration of R. This coherence transfer into population is the key mechanism of the stimulated photon-echo protocol, where optical population becomes a critical parameter for photon storage[3]. Here, the spectral grating is a population modulation in the optical inhomogeneous broadening regardless of data (D) pulse intensity, because the density matrix approach is quantum mechanical[26].

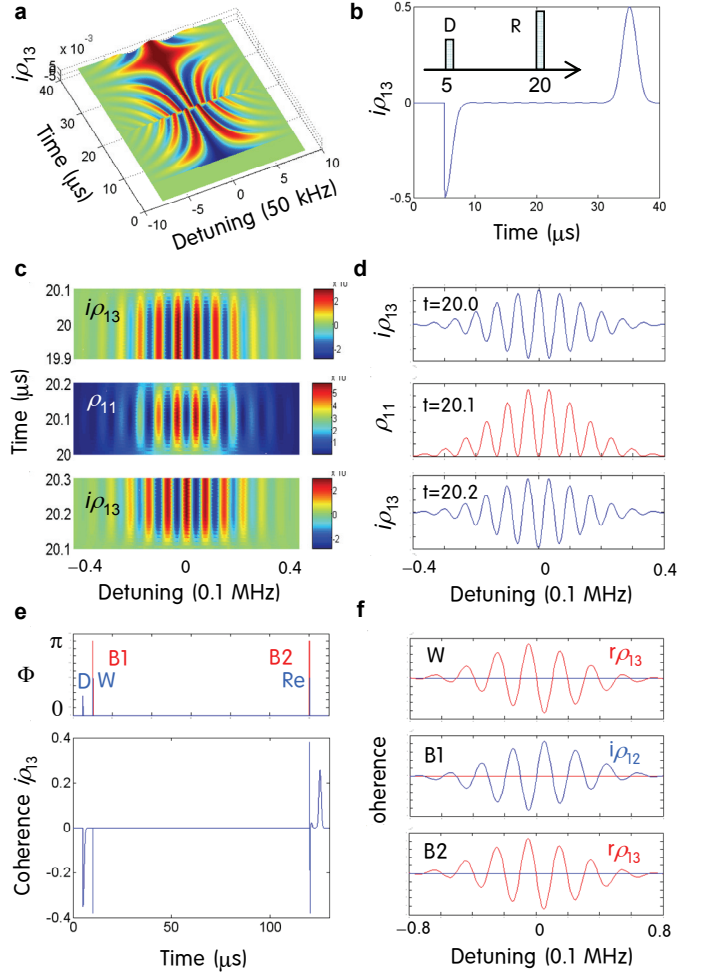

**Figure 2| Coherence conversion between $Im\rho_{13}$ and $\rho_{11}$ ($\rho_{33}$) in a two-level system composed of ground state |1> and excited state |3>. a, b,** Absorption ($Im\rho_{13}$) as functions of time and detuning for two-pulse photon echoes. Inset: pulse sequence. The pulse area of D and R is $\pi/2$ and $\pi$, respectively. **c, d,** Coherence conversion by a rephasing pulse R in (**b**) between $Im\rho_{13}$ and $\rho_{11}$. Optical inhomogeneous broadening Δ=340 kHz (FWHM). **e, f,** Optically locked echoes. **e,** Top panel: pulse sequence of Fig. 1b. **f**, Red: $Re\rho_{13}$, Blue: $Im\rho_{12}$. Pulse area of D is $\pi/4$. Pulse area of W and Re is $\pi/2$. Pulse area of B1 and B2 is $\pi$ and $3\pi$, respectively. Optical inhomogeneous broadening, Δ=680 kHz (FWHM). $\Gamma_{31}=\Gamma_{32}=\gamma_{31}=\gamma_{32}=5$ kHz; $\Gamma_{12}=\gamma_{12}=0$.

Figures 2e and 2f present the optically locked stimulated photon echoes achieved by dividing R into two halves, W and Re, and inserting the optical locking pulses B1 and B2 between them (as shown in Fig. 1). Figure 2e shows an optically locked echo simulation, where optical population decay-caused coherence loss is prohibited: The optical decay time $T_1$ is $T_1=1/[2\pi(\Gamma_{31}+\Gamma_{32})]=16.7$ μs. Coherence transfers between $Re\rho_{13}$ and $Im\rho_{12}$ due to the population transfer by B1 and B2 are shown in Fig. 2f. This coherence transfer is due to exact population transfer from state |3> to state |2>. Thus, optical population decay-dependent decoherence is simply replaced by ultraslow spin population decay-dependent decoherence, which is the fundamental mechanism of the optical locking for ultralong photon storage. Just as the stimulated photon echo is independent of optical phase decay rate[3,27], so the optically locked echo is also independent of the spin phase decay rate. It is well known that the Bloch vector model that explains the phase decoherence process very well in two-pulse photon echoes is not appropriate to the stimulated (three pulse) photon echoes.

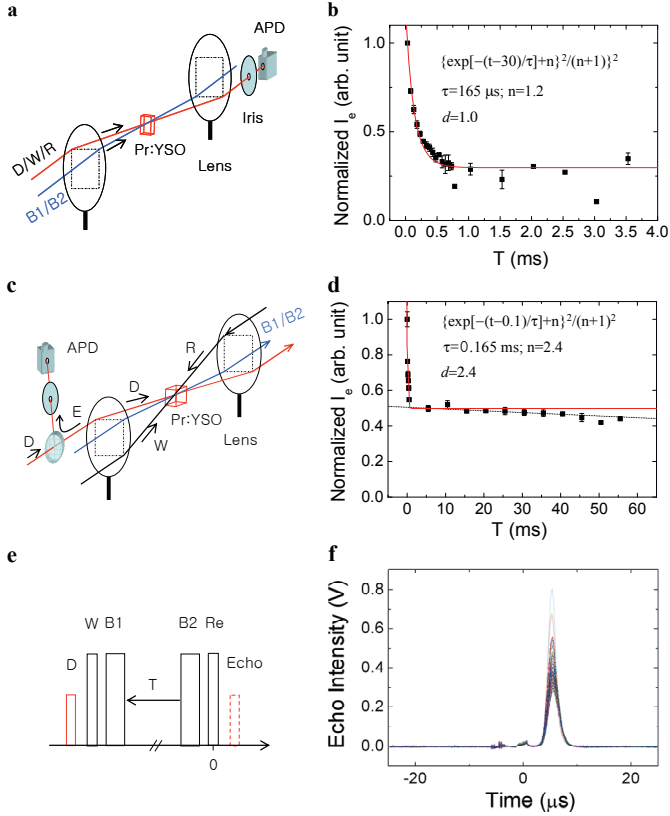

**Figure 3| Observations of ultralong optically locked echoes. a,** Schematic of forward propagation. **b,** Optically locked echoes with B1 and B2 as a function of delay T (in Fig. 1). The delays of R from B2 and W from D are fixed at 2 μs. Red lines are for $\{\exp[-(t-\Delta T)/\tau]+n\}^2/(n+1)^2$, where $\Delta T$ is the minimum delay and n≥0. The pulse duration of D, W, and R in Fig. 1 is set at 1 μs for the π/2 pulse area and 4.6 mW in power. The pulse duration of control pulse B1 (B2) is set at 1.2 μs (3.6 μs) for the π (3π) pulse area at 14.5 mW in power. The optimum power of the light B1 and B2 is predetermined by Rabi flopping measurement. An avalanche photodiode APD1 detects light in the line of D, W, and Re including conventional photon echoes, and the detected signals are directly fed into a digital oscilloscope and recorded by averaging 10 samples. The spot diameter [exp(−2) in intensity] of B1, B2, D, W, and Re are ~300 μm. The Pr:YSO sample in a liquid helium cryostat (Advanced Research System) is kept at a temperature of ~5 K. The resultant absorption of a data pulse D is ~70%, where optical depth d (d=$\alpha l$, where $l$=1 mm) is 1.0. All light pulses are vertically polarized, and propagate along the crystal axis of the medium (Pr:YSO). The angle between beams D and B1 is 12.5 miliradians overlapping each other by 90% through the sample ($Pr^{3+}$:$Y_2SiO_5$) of 1 mm in length. The dashed green line indicates 50% echo efficiency in amplitude. **c,** Schematic of backward propagation. **d,** Optically locked echoes with B1 and B2 as a function of delay T for (**c**). The delays of Re from B2, and W from D, are fixed at 2 μs. The angle between beams D and B1 is 12.5 milliradians and overlapped by 80% through the sample ($Pr^{3+}$:$Y_2SiO_5$) of 3 mm in length. The resultant absorption of data pulse D is 90%, where optical depth d (d=$\alpha l$, where $l$=3 mm) is 2.4. Each spot diameter [exp(−2) in intensity] of B1 and B2 is 340 μm. The spot diameter [exp(−2) in intensity] of D, W, and Re is 280, 390, and 590 μm, respectively. The peak power of D, W, Re, B1 and B2 is 7, 23, 10, 12, and 12 mW, respectively. The pulse duration of D, W, Re, B1, and B2 is 0.76, 0.7, 2, 1.2, and 3.6 μs, respectively. The Pr:YSO sample in a liquid helium cryostat (Advanced Research System) is kept at a temperature of ~6 K. The dotted line is for τ=2 sec, which is $T_2^{spin}$ at ~6 K (see Fig. 9 of Ref. 29). **e,** Pulse sequence for (**f**). **f,** All measured echoes in (**d**) are overlapped for comparison.

Detailed optically locked echo experiments are shown in Fig. 3 for both forward and backward propagation schemes at different optical depths. The purpose of the backward propagation scheme[4,6] is to solve the intrinsic echo reabsoprtion problem inevitable in a forward scheme by letting echoes retrace backward along the data trajectory[4,28]. The purpose of using two different optical depths is to demonstrate the imperfect population transfer-caused coherence loss in a dilute medium[25]. Since B1 and B2 function optical deshelving of the phase locked atoms by W from state |3> to state |2>, the phase matching condition is the same as for the conventional stimulated photon echoes[3]:

$$\vec{k}_E = -\vec{k}_D + \vec{k}_W + \vec{k}_R, \quad (3)$$
$$\varpi_E = -\varpi_D + \varpi_W + \varpi_R, \quad (4)$$

where $\vec{k}_i$ and $\varpi_i$ are wave vector and angular frequency of the pulse *i*, respectively. Unlike the optical locking applied to the rephasing process[1,11,23], optical locking applied to the spectral grating in Fig. 1 freezes the phase decay process of individual atoms from both optical and spin dephasing. Thus, a spin dephasing-independent photon storage protocol should be expected, where ultralong storage time extension via a perfectly collective coherence swapping between optical and spin transitions is possible even in a spin jitter-based noisy optical system. Note here that the photon echo methods[1-12] have been accepted as quantum memory protocols because, in a weak field, a classical data pulse can be treated quantum mechanically[2,29].

Figure 3a presents a forward propagation scheme, and Fig. 3b shows the results as a function of delay time T (B2 delay from B1) in a lower optical depth d (d=$\alpha l$=1.0; $l$=1 mm). Due to echo reabsorption, the measured retrieval efficiency (ECHO intensity ratio to the DATA intensity at T=0) is about 0.2%. The decay time τ (τ=165 μs) of the observed optically locked echoes, however, corresponds to the optical population decay time $T_1^{opt}$. Although the decay time is much longer than the spin dephasing time $(T_2^{spin})^*$, it is much shorter than spin population decay time $T_1^{spin}$: $T_1^{opt}$ ~ 165 μs (ref. 32; see also Fig. S2 in Supplementary Information); $(T_2^{spin})^*$ ~ 10 μs (refs. 11,23; $T_1^{spin}$~100 s at 1.4 K (ref. 30). This demonstrates that the mechanism of optical locking applied to the spectral grating in the present paper is different from the rephasing halt in refs. 11 & 23. The optical population decay-dependent decoherence is due to the remnant population in state |3>, where the coherence decay must be governed by the optical decay time $T_1^{opt}$ (will be discussed below). The equation in Fig. 3b given to the decay curve will be discussed in Fig. 4.

Figure 3c illustrates a backward propagation (phase conjugate) scheme, and Fig. 3d shows the results for optically locked echoes as a function of delay T in a greater optical depth ($d=\alpha l=2.4$) using a longer sample ($l=3$ mm). The decay time of the measured echoes is the same as in Fig. 3b. However, the saturation level of the echo intensity increases from 30% in Fig. 3b to 50% due to fewer remnant atoms in the excited state |3>. Here, the 50% echo efficiency indicates a no-cloning regime without post-selection, where quantum error correction protocols can be applied[31]. Compared with Fig. 3b, the higher echo saturation level in Fig. 3d is a direct proof of the remnant population-caused coherence leakage[25]. Thus, in an optically dense medium, where nearly perfect population transfer occurs, the decay curve should follow the spin population decay time without the $T_1^{opt}$-dependent sudden drop in echo intensity (will be discussed in Fig. 4).

Owing to the backward propagation scheme as proposed in ref. 4, echo retrieval efficiency at T=0 is measured at 10%, which is 50 times higher than in Fig. 3b (see Fig. S3 in Supplementary Information). This echo enhancement presents the first experimental proof of echo reabsorption cancellation in a backward scheme proposed in refs. 4 and 6. Thus, nearly 100% retrieval efficiency can be obtained in an optically dense medium if the beams are nearly perfectly overlapped inside the medium. Here, a stimulated photon echo decay curve should follow $T_1^{opt}$ if the delay $T_W-T_D$ is too small compared with the optical phase decay time $T_2^{opt}$ ($T_2^{opt} \sim 110$ μs; see ref. 32), otherwise shortens[27]. In Fig. 3, the delay of WRITE pulse W from the DATA pulse D, $T_W-T_D$, is just 2 μs. Therefore, the optically locked echoes must decay according to the spin population decay time. The dotted line is for $\tau = 2$ sec, which indicates the spin population decay time at 6 K (see Fig. 9 of ref. 30).

Figure 3f shows all optically locked echo signals of Fig. 3d overlapped at one position. For each data point in Fig. 3d, three sets of echoes are measured and averaged. As shown in Fig. 3f, the echo intensity becomes decreased and saturated as the delay time T increases. The saturation level of echo intensity should depend on optical density of the medium (will be discussed in Fig. 4).

To support the analysis of imperfect population transfer-caused (or remnant population-caused) coherence loss discussed in Fig. 3, numerical simulations of optically locked echoes with improper optical locking pulses, B1 and B2, are shown in Fig. 4. To satisfy the remnant atom-caused coherence loss, the following equation is used for the decay curve:

$$y(t) = \{\exp[-(t-\Delta T)/\tau] + n\}^2 / (n+1)^2, \quad (5)$$

where $\Delta T$ is the shortest T for the first echo data, $\tau$ is $T_1^{opt}$, and $n$ ($n \geq 0$) is determined by the remnant atom-caused coherence leakage. Equation (5) is intuitively obtained from the analysis of population transfer as a function of pulse area (see Fig. S4 in Supplementary Information). In comparison, the pulse area of B1 varies for a fixed B2 ($\Phi_{B2}=2.4 \pi$). For simplification, spin decay rates are set at zero, resulting in no decay if a perfect population transfer is achieved by B1 and B2. The dotted curves refer to equation (5). As shown in Figs. 4a~4d, echo retrieval efficiency increases, as coherence leakage (due to remnant atoms) decreases. For all cases in Fig. 4, the echo decay rate is the same as the population decay rate $\Gamma_3$ ($\Gamma_3=\Gamma_{31}+\Gamma_{32}$) supporting the sudden coherence drop in Fig. 3 is due to the imperfect population transfer in a dilute sample. With the same optical depth, the coherence recovery in Figs. 4b (88% in amplitude) and 4d (46% in amplitude) corresponds to Figs. 3d (50% in intensity; 70% in amplitude) and 3b (30% in intensity; 55% in amplitude), respectively (see Fig. S4 and Table S1 in Supplementary Information). Thus, to satisfy zero coherence leakage of the present optically locked echoes, use of an optically dense medium is an essential requirement.

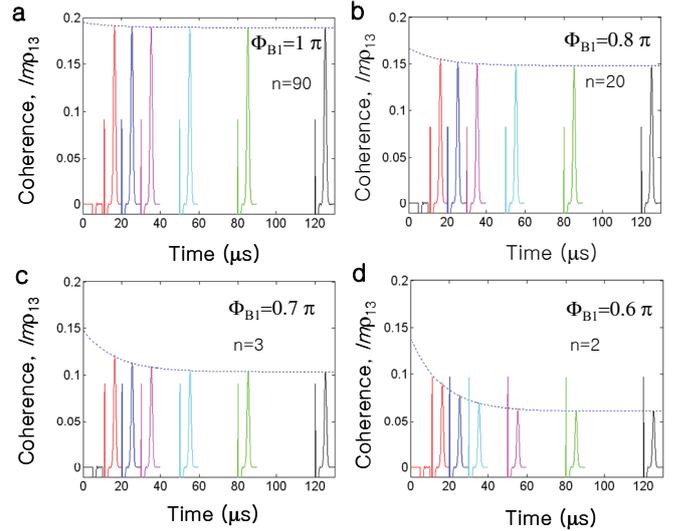

**Figure 4| Numerical simulations of the remnant population-caused coherence leakage. a~d,** Delay T of B2 from B1 is set at 0.9 (red), 10 (blue), 20 (magenta), 40 (cyan), 70 (green), and 110 μs (black). The positions of light pulses D, W, and B1 are $T_D=5$, $T_W=10$, $T_{B1}=10.1$ μs, respectively. The pulse duration of D, W, B1, and R is 0.1 μs. The pulse duration of B3 is 0.3 μs. The pulses R and B2 move together, where their separation is 0.1 μs. The pulse area of D is $\pi/4$. All resultant echoes are superposed. The pulse area $\Phi_{B2}$ of B2 is fixed at $\Phi_{B2}=0.8 \pi$, while $\Phi_{B1}$ varies to test the coherence loss due to imperfect population transfer (or remnant atom-caused coherence loss) by B1 pulse excitation. The n is for Eq. 5. The optical inhomogeneous width (FWHM) is 680 kHz. $\Gamma_{31}=\Gamma_{32}=\gamma_{31}=\gamma_{32}=5$ kHz.

In conclusion, an on-demand photon storage protocol for ultalong quantum memories was proposed and experimentally demonstrated using optical locking in stimulated photon echoes, where the storage time far exceeds the critical constraint of spin phase decay time, but is limited by spin population decay time in the order of seconds. Moreover, the inevitable echo reabsorption problem in photon echoes was solved by adapting a phase conjugate scheme. The calculated population inversion-caused spontaneous emission noise in photon echoes is negligibly small to alter the quantum fidelity in quantum memories. Thus, the present optically locked echo protocol holds potential for long-distance quantum communications with ultralong photon storage time and near perfect retrieval efficiency.

**METHOD SUMMARY**

**Experimental methods.** The three laser beams in Fig. 1 are the modulated output of a ring-dye laser (Tecknoscan) pumped by a 532 nm laser (Coherent Verdi). Relative frequency adjustment for $\omega_0$, $\omega_1$, and $\omega_2$ is achieved by using acoustooptic modulators (Isomet) driven by radio frequency (rf) synthesizers (PTS 250). The data pulse D is made by an arbitrary waveform generator (AWG610) for a hyperbolic secant pulse: $\Omega(t) = \Omega_0 \sec h[\beta(t-t_0)]$, where $\Omega_0$ is the maximum Rabi frequency of the pulse and $\beta$ is real to determine the pulse duration. Other pulses are square pulses, so that a perfect population transfer by B1 and B2 cannot be made. Duration of each pulse of the laser beam is controlled by a rf switch (Minicircuits) and a digital delay generator (SRS DG 535). The repetition rate of the light pulse train is 10 Hz. The lights P, D, W, B1, B2, and Re are generated from the same laser beam. The light pulse P ($\omega_0;\omega_2$) is used for initial preparation of ground state population redistribution: $\rho_{11}=1$; $\rho_{22}=0$; $\rho_{33}=0$. The control lights B1 and B2 for optical locking are copropagating. The angle between D and B1 is 12.5 milliradians with overlap of ~90% (~80%) along the 1 mm (3 mm) sample. The measured optical phase decay time ($T_2$) is 25 μs at ~5 K (see Ref. 23). Within the allowed bandwidth given by modified optical inhomogeneous broadening, the pulse area is simply adjusted according to the pulse duration. An avalanche photodiode (Hamamatsu APD) detects light signals. The APD-captured signals are directly fed into a digital oscilloscope and recorded by averaging 10 samples. The light B1 and B2 is predetermined for the pulse area of $\pi$ and $3\pi$, respectively, unless otherwise indicated. The Pr:YSO sample in a liquid helium cryostat (Advanced Research System) is kept at a temperature of 5~6 K. The effective atom broadening due to the redistribution procedure is determined by laser jitter at ~300 kHz. The resultant absorption of a data pulse D is ~70% for 1 mm sample in Fig. 2 and ~92% for 3 mm sample in Fig. 3 (see also Fig. S3 of Supplementary Information). All light pulses are vertically polarized, and propagate along the crystal axis of the medium (Pr:YSO). Here a significant difference occurs between temperatures 4.2 K and 5 K, because not only optical phase decay but also spin population decay heavily depends on temperature (see Ref. 29). For example, at 8 K, spin $T_1$-based spectral hole burning disappears, where the optical phase decay rate exceeds spin hyperfine splitting of 27 MHz: At 1.4 K, the optical phase decay rate is 2.4 kHz. Thus, the echo efficiency significantly drops as temperature increases at above 4.2 K, even with one degree increase. Liquid helium consumption rate at 5 K decreases by eight times compared with that at 2 K.

**Numerical calculations.** For the numerical calculations of Figs. 2 & 4, time-dependent density matrix equations are numerically solved under the rotating wave approximation for the three-level system of Fig. 1 composed of states |1>, |2>, and |3>. The density matrix approach works powerfully with an ensemble system interacting with coherent laser fields owing to statistical information as well as quantum mechanical information. The equation of motion of the density matrix operator $\rho$ is determined from Schrödinger's equation[26]:

$$\frac{d\rho}{dt} = -\frac{i}{\hbar}[H,\rho] - \frac{1}{2}\{\Gamma,\rho\}, \qquad (6)$$

where $\{\Gamma,\rho\}$ is $\Gamma\rho + \rho\Gamma$, $H$ is the Hamiltonian, $\hbar$ is the Planck's constant divided by $2\pi$, and $\Gamma$ is decay rate. The density operator $\rho$ is defined by $\rho = |\Psi\rangle\langle\Psi|$, where $|\Psi\rangle$ is the state vector. By solving equation (6) the following is obtained as a main coupled equation[23]:

$$\frac{d\rho_{13}}{dt} = -i\Omega_1(\rho_{11}-\rho_{33}) - i\Omega_2\rho_{12} - i\delta_1\rho_{13} - \gamma_{13}\rho_{13}, \quad (7)$$

where $\Omega_1$ and $\Omega_2$ are Rabi frequencies of light at $\omega_1$ and $\omega_2$, respectively. The term $\gamma_{13}$ is the phase decay rate between states |1> and |3>. For the numerical calculations, nine time-dependent density matrix equations are solved under rotating wave approximations. Gaussian shaped optical inhomogeneous broadening of 340 kHz (680 kHz for Fig. 4) is set at full width at half maximum for the optical transitions. For the calculations, the 0.8 MHz (1.6 MHz for Fig. 4) inhomogeneous region is divided into 401 segments.

**Supplementary Information** is linked to the online version of the paper at www.nature.com/nature.

**Acknowledgmens** This work was supported by the CRI program (No. 2010-0000690) of the Korean government (MEST) via National Research Foundation. BSH acknowledges that J. Hahn contributed to the experiments. BSH thanks to S. Kröll of Lund University, J. Howell of University of Rochester, and M. D. Lukin of Harvard University for helpful discussions.

# Supplementary Information
## 1. Review of spontaneous emission-caused quantum noise in photon echoes

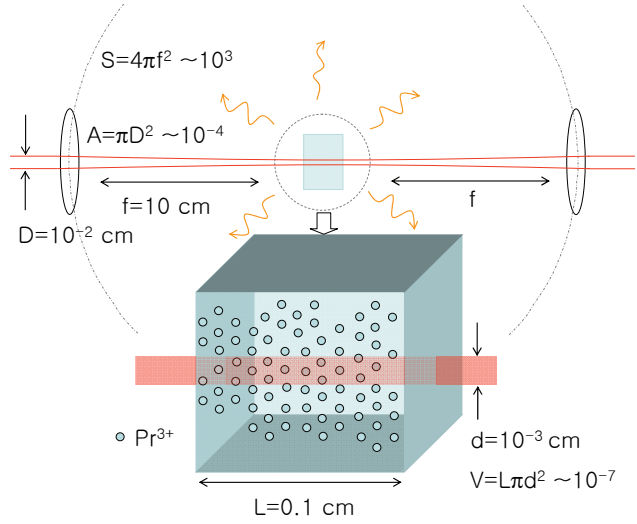

Fig. S1. Schematic diagram of light propagation through a Pr:YSO sample

Recently N. Sangouard et al discussed a spontaneous emission noise problem in three-pulse photon echoes for quantum memory applications [see PRA 81, 062333 (2010)]. Although rephasing pulse-excited atoms cause spontaneous emission noise affecting the fidelity of photon echoes, the discussion in PRA 81, 062333 (2010) was severely exaggerated for practical situations neglecting echo directionality in a pencil-like geometry, echo time window, and effective atom numbers that contribute to the noise. Contrary to the conclusion in PRA 81, 062333 (2010), the spontaneous emission-cause quantum noise on the photon echo signal poses a negligibly small threat to the fidelity of quantum memories. Detailed calculations with practical parameters are followed. For a rare earth $Pr^{3+}$ (0.05 at. %) doped $Y_2SiO_5$ (Pr:YSO), the $Pr^{3+}$-ion number $N_0$ in a unit volume is $N_0=4.7 \times 10^{18}$ cm$^{-3}$ [see B. A. Maksimov et al., Sov. Phys.-Doklady 13, 1188 (1969)]. The interaction volume V of Pr:YSO with single photon light focused by a 10 cm focal length lens is V~$10^{-7}$ cm$^3$ for a medium's length L (L=1 mm) and Gaussian width d (d=10 μm) at a focal point. For this volume V, the calculated effective atom number N is $N=VN_0 \sim 5 \times 10^{11}$. The ratio η of a minimum single photon pulse duration ΔT (ΔT=0.1 ns) for the 4 GHz inhomogeneous width of Pr:YSO to the spontaneous emission decay time $T_1$ ($T_1$=160 μs) is $\eta = \Delta T/T_1 \sim 6 \times 10^{-7}$. Thus, for the case of full inversion, the noise-contributed atom (or photon) number $N_e$ to the echo signal is $N_e = \eta N = 3 \times 10^5$. Considering the data pulse area A on the virtual spontaneous emission sphere S made by the 10 cm focal length lens in the pencil-like photon echo geometry, the echo area ratio α is $\alpha = A/S = 10^{-7}$. Therefore, the final effective atom number $N_f$ contributing to the quantum noise by the spontaneous emission decay process is $N_f = \alpha N_e = \alpha \eta V N_0 \sim 0.03$ resulting in no more than 10% fidelity loss in the retrieval efficiency of the photon echoes. Therefore, the discussion regarding spontaneous emission noise to the echo fidelity in Ref. xx for (three-pulse) photon echoes is severely exaggerated and overlooked practical conditions. Moreover, with use of multiphoton entangled light or even squeezed light the spontaneous emission noise becomes nearly negligible.

## 2. Measurement of three-pulse photon echoes without B1 and B2

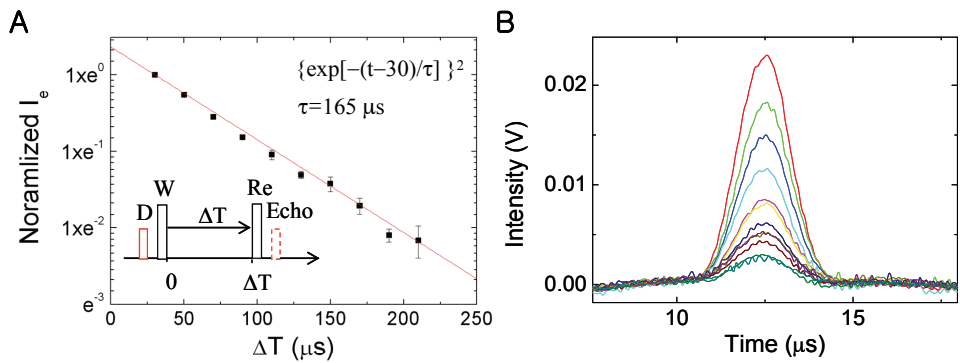

Fig. S2. Conventional three-pulse photon echoes. (A) Measured echo intensity versus delay time ΔT. (B) Overlapped echo signals of (A).

## 3. Measurement of optically locked echoes in a phase conjugate scheme with B1 and B2

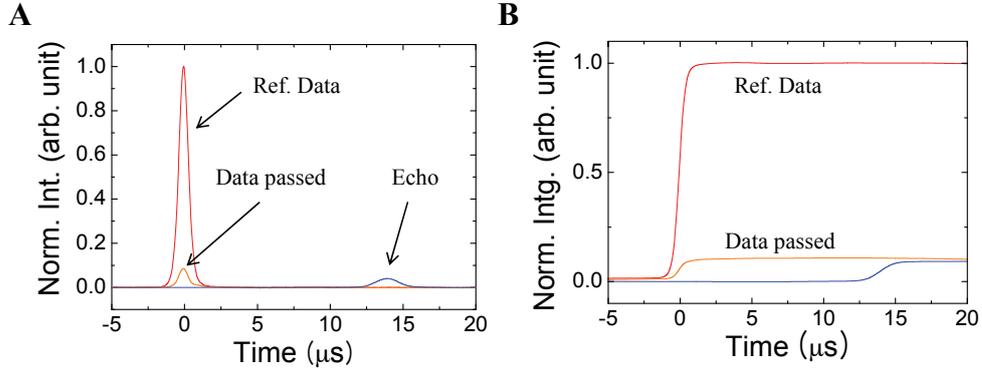

Fig. S3. (A) Measured light pulse intensity by APD. (B) Each integrated light pulse area of (A). The scheme is for a backward propagation of the optically locked echo in Fig. 3d. The measured echo intensity ratio in (B) is ~10% of the original data pulse intensity.

## 4. Analysis of imperfect population transfer by B1 and B2

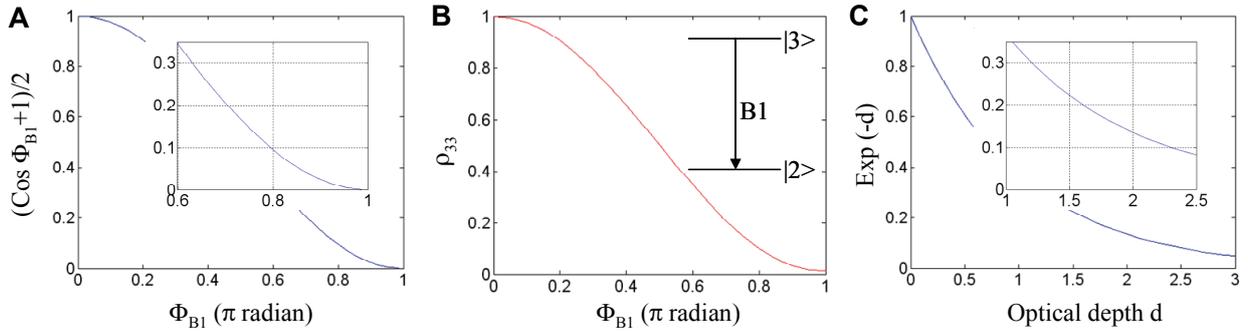

Fig. S4. (A) cos $\Phi_{B1}$ versus $\Phi_{B1}$. (B) Excited state population $\rho_{33}$ versus B1 pulse area $\Phi_{B1}$. Initial condition: $\rho_{33}$=1; $\rho_{22}$=0. (C) Absorption versus optical depth.

In Fig. S4, the excited state population as a function of B1 pulse area in a perfectly absorbing medium is compared with the optical depth-dependent absorption ratio due to Beer's law. The population left in state |3> is comparable with a cosine function (see A and B). In (C), for d=1, nonabsorbed (or remnant) atom ratio is ~40% after B1 pulse excitation, which is comparable with 0.6π pulse area of B1 in (A). For d=2.4, the remnant atom ratio is ~10%, which is 0.8π in pulse area of B1 in (A). For $\Phi_{B1}$=0.6π in Fig. 4d ($\Phi_{B1}$=0.8π in Fig. 4b), the echo amplitude ratio is 0.06/0.13~46% (0.15/0.17~88%). Because the optical depth used in Fig. 3b (Fig. 3d) is d=1.0 (d=2.4), the corresponding comparison is Fig. 3b vs. Fig. 4d (Fig. 3d vs. Fig. 4b), the observed echo intensity efficiency at saturation level in Fig. 3b (Fig. 3d) is 30% (50%). Corresponding amplitude ratio is comparable with that in Fig. S4. Therefore, the analysis of remnant atom-caused coherence loss in Fig. 3 is supported.

Table S1. Optical depth versus pulse area of B1 in Fig. S4.

| Optical depth d | 1.0 | 2.4 |
| --- | --- | --- |
| Untransferred $\rho_{33}$ | 40% | 10% |
| $\Phi_{B1}$ (π) | 0.6 | 0.8 |